# Preparation of pseudopure state in a cluster of dipolar-coupled spins with "unresolved" spectrum

Jae-Seung Lee and A. K. Khitrin

*Department of Chemistry, Kent State University, Kent, OH 44242-0001*

**Abstract**

A method of creating pseudopure spin states in large clusters of coupled spins is described. It is based on filtering multiple-quantum coherence of the highest order followed by a time-reversal period and partial saturation. Experimental demonstration is presented for a cluster of six dipolar-coupled proton spins of a benzene molecule in liquid crystalline matrix, and the details of spin dynamics are studied numerically.

03.67.Lx, 82.56.-b

## 1. INTORDUCTION

At present, the largest systems for experimental quantum information processing (QIP) can be explored with nuclear magnetic resonance (NMR) [1,2]. With liquid-state NMR, quantum algorithms for up to seven quantum bits (spins $1/2$) have been implemented [3]. The first step of quantum algorithms is usually the state initialization or preparation of a pseudopure state [1,2]. Conveniently, the algorithms start with a pseudopure ground state with equal populations of all states except the ground state. Until now, all practical schemes used transition-selective or qubit-selective pulses for state initialization. Therefore, the size of a spin system which can be prepared in pseudopure state has been limited by spectral resolution. For an $N$-spin system, the number of quantum levels $2^N$ grows very fast with increasing number of spins, and the number of peaks in NMR spectrum increases even faster. As an example, the maximum number of peaks for $N$ dipolar-coupled spins is $\left(\dfrac{2N}{N+1}\right) \sim 2^{2N}$. The number of operations (length of a pulse sequence), needed for state preparation also grows exponentially with the number of spins. With multi-frequency irradiation, it is possible to arrange a more efficient simultaneous evolution of all populations to desired values [4]. However, this approach also requires a well-resolved spectrum.

Compared to the thermal equilibrium state, which contains all quantum states, the highest-order multiple-quantum (HOMQ) coherence is built of only two states, $|u><d|$ and $|d><u|$, where $|u>$ is the state with all spins up and $|d>$ is the state with all spins down. The relation between the HOMQ coherence and the maximally entangled "cat" state $(|u>+|d>)(<u|+<d|)$ has been discussed in [5], where the HOMQ coherence has been used to build a pseudopure ground state of a six-qubit subsystem in a system of seven qubits. The implemented logic circuit used qubit-selective pulses and required a resolved spectrum. At the same time, there exist efficient techniques for exciting MQ coherences [6-8], which use sequences of non-selective hard pulses and can be applied to systems with unresolved spectra.

Here we describe a scheme for creating pseudopure states, which does not require a resolved equilibrium spectrum and can be applied to larger and more complex spin systems.

## 2. EXPERIMENT

The experiment consists of the following steps: (1) excitation of multiple-quantum coherences (preparation period), (2) filtering the HOMQ coherence, (3) time-reversal period, and (4) partial saturation. There are many pulse sequences exciting multiple-quantum coherences during the preparation period. For a dipolar-coupled spin system, the multi-pulse sequence with an eight-pulse cycle [6] is known to be very efficient. It creates the double-quantum effective Hamiltonian

$$H_{av} = H^{+2} + H^{-2} = -\frac{1}{2}\sum_{i<j} D_{ij}(I_{i+}I_{j+} + I_{i-}I_{j-}), \qquad (1)$$

where $I_{\pm} = I_x \pm iI_y$ and $D_{ij}$'s are the dipolar coupling constants. This effective Hamiltonian excites MQ coherences of even order. The HOMQ coherence can be excited in clusters of $2+4n$ ( $n=0,1,2,\ldots$ ) coupled spins. In finite clusters, intensities of individual MQ coherences oscillate. Total duration of the pulse sequence is adjusted to correspond to one of the maxima of HOMQ coherence. A combination of phase cycling and 180° pulse is used to average out all MQ coherences except the HOMQ coherence [9]. The deviation density matrix after this temporal averaging is $i(|u\rangle\langle d| - |d\rangle\langle u|)$ (Its trace is zero and, therefore, it can not be directly converted by a unitary transformation into a pseudopure state). Then the time-reversal period of the same duration follows. The pulse sequence for this period is the same except that the phases of all pulses are shifted by 90°. This sequence creates the effective Hamiltonian $-H_{av}$. After the time-reversal period, off-diagonal elements of the density matrix are small if the duration of the preparation period corresponds to a maximum of HOMQ intensity. Then, from conservation of $\text{Tr}\{r\}$, $\text{Tr}\{r^2\}$, and $\text{Tr}\{r^2 H_{av}^2\}$ during the unitary time-reversal period we conclude that the deviation density matrix becomes $|u\rangle\langle u| - |d\rangle\langle d|$. Our computer simulation of the dynamics of this system showed that when the preparation time corresponds to one of the major maxima of the HOMQ intensity, the HOMQ coherence is practically ideally converted into the state $|u\rangle\langle u| - |d\rangle\langle d|$ after the time reversal period of the same duration. The details of this simulation for a six-spin ring system will be discussed later.

The state $|u\rangle\langle u| - |d\rangle\langle d|$ is still traceless, which means that a non-unitary operation is needed to prepare the pseudopure ground state $|u\rangle\langle u|$. We used partial saturation, which redistributes the excessive population of the state $|d\rangle\langle d|$ among other states but does not change the population of the state $|u\rangle\langle u|$. Irradiation with wide spectrum causes single-quantum transitions between states and equalizes the populations. If this irradiation does not contain frequencies of transition from the ground state, the population of the ground state remains "trapped". It is not necessary that the spectrum of irradiation includes all the transition frequencies. Most of the states are connected, by single-quantum transitions, to many other states and the excessive population of the state $|d\rangle\langle d|$ would leak to other states even if not all single-quantum transitions are excited. In our experiments we used shaped Gaussian pulses for this type of partial saturation.

Linear-response spectra corresponding to pseudopure states can be useful for the state identification because they contain smaller number of peaks than equilibrium spectra. As an example, for the pseudopure ground state, the linear-response spectrum has no more than $N$ peaks. It can be well resolved and used for the state identification in very large systems of coupled spins.

The physical system we used was six dipolar-coupled proton spins of a benzene molecule dissolved in liquid crystal ZLI 1167. Details of the sample preparation and parameters of spin Hamiltonian are described elsewhere [10]. The experiments have been performed with a Varian Unity/Inova 500 MHz NMR spectrometer. The pulse sequence for preparation of the pseudopure ground state $|u\rangle\langle u|$ is displayed in Fig. 1. The 90° pulse duration was $4.46\mu s$, the intervals between pulses were $\Delta = 5.54\mu s$, $\Delta' = 15.54\mu s$, and the number of cycles $n$ for both preparation and time-reversal periods was 20. Each of the gradient pulses $G$ had duration of 2 ms and amplitude 1.25 Gauss/cm. The frequency of the two shaped Gaussian pulses was centered at the frequency of the single-quantum transition from the state $|d\rangle\langle d|$. Both of the shaped pulses had 1.5 ms duration and maximum amplitude $\gamma B_1/2\pi = 1.5$ kHz. The pulse R is a small-angle reading pulse with a flip angle 5°.

The spectrum at thermal equilibrium is shown in Fig. 2(a). Positions and intensities of 76 peaks are determined by dipole-dipole interactions between six proton spins of the benzene molecule [11]. For this comparatively small system the spectrum is well resolved. However, it is virtually impossible to use selective pulses for manipulating the states, because the length of a single selective pulse would be comparable to spin-lattice relaxation time $T_1$, which is 3.3 s in this system.

The linear-response spectrum corresponding to the state $|u><u|-|d><d|$ is presented in Fig. 2(b). For this state, the deviation density matrix consists of only two elements, one positive and one negative. Due to the high symmetry of a benzene molecule, there is only one allowed single-quantum transition from each of the two states, $|u><u|$ and $|d><d|$. The spectrum was obtained with a small-angle reading pulse which satisfies a linear-response condition and gives intensities of the peaks proportional to the differences of the populations. A large-angle pulse produces for this state a spectrum which includes single-quantum transitions between all levels with the total spin $S = 3$. Our simulation of the dynamics of this spin system with the effective Hamiltonian $H_{av} = H^{+2} + H^{-2}$ showed that up to 15% of the initial intensity can be converted into six-quantum coherence. The total intensity of the two peaks in Fig. 2(a) is about 11% of integral intensity of the thermal equilibrium spectrum. The peaks are about 1.8 times higher than corresponding peaks in equilibrium spectrum. Since we used a small-angle reading pulse to identify the states, we used the same small-angle pulse for exciting the equilibrium spectrum, so that the peaks' intensities could be compared.

The central frequency of the saturating Gaussian pulses was set to the frequency of single-quantum transition from the state $|d><d|$ (the left peak in Fig. 2(b)). The spectrum of the saturating pulse covers mostly the left side of the spectrum while the intensity at the frequency of transition from the state $|u><u|$ (the right peak in Fig. 2(b)) is very small. As one can see in Fig. 2(c), population of the state $|u><u|$ is practically unaffected by the saturating pulses. The linear-response spectrum in Fig. 2(c) verifies that the pseudopure ground state with the deviation density matrix $|u><u|$ has been created. It should be noted that the spectra in Figs. 2(a) and 2(b) are not multiple-quantum spectra. They are conventional single-quantum linear-response spectra for unusual and strongly correlated spin states.

## 3. SIMULATION

The computational power of a quantum computer (QC) comes from high dimensionality of the corresponding Hilbert space, which increases exponentially with the number of qubits. This makes a classical computer inefficient in simulating a quantum one [12]. At present, the experimental state-of-the-art is the implementation of the quantum factoring algorithm with a seven-qubit system, using a liquid-state NMR technique [3]. The dynamics of a system of this size is too complicated for analytical description. At the same time, experimentally realizable QCs are still small enough to be easily simulated with classical computers. Today, a supercomputer can simulate quantum systems with up to 15 qubits [13]. This means that a simulation with a classical computer remains a useful tool at the current level of progress in experimental quantum information processing (QIP).

Experimental realization is affected by various imperfections which are hard to eliminate, control or take into account. The goal of this simulation was to find theoretical efficiencies of all steps of the density matrix conversion and to see how precise, in the ideal situation, can be a conversion of the HOMQ coherence into the diagonal state $|u><u|-|d><d|$. The simulation has been performed with a PC using the *Mathematica* package. Since the experimental realization has been demonstrated for a system of six dipolar-coupled proton spins of a benzene molecule in liquid-crystalline matrix, we have chosen a system of six spins $1/2$, arranged as a regular hexagon. Therefore, the ratios of the dipole-dipole coupling constants are given by

$$D_{12} : D_{13} : D_{14} = 1 : \frac{1}{3\sqrt{3}} : \frac{1}{8}, \qquad (2)$$

where $D_{12}$, $D_{13}$, and $D_{14}$ are the dipolar coupling constants between nearest neighbors, next nearest neighbors, and spins at opposite sites, respectively. The density matrix at time $t$ after evolution with the Hamiltonian of Eq. (1) is given by

$$\boldsymbol{r}(t) = \exp(-iH_{av}t)\,\boldsymbol{r}(0)\exp(iH_{av}t). \qquad (3)$$

For our scheme of the pseudopure state preparation, two cases are of particular interest: when the initial density matrix is the thermal equilibrium state, $\boldsymbol{r}(0) = I_z$, and when it is the six-quantum MQ coherence, $\boldsymbol{r}(0) = i(|u\rangle\langle d| - |d\rangle\langle u|)$. After calculating the evolution (Eq. (3)), we grouped the terms in $\boldsymbol{r}(t)$ according to their MQ order $n$: $\boldsymbol{r}(t) = \sum_n \boldsymbol{r}_n$, and used $\mathrm{Tr}\{\boldsymbol{r}_n^2\}$ as a convenient measure of intensities of the MQ coherences of different orders. It should be noted that the evolution of the MQ coherences in a six-spin ring, with the Hamiltonian of Eq. (1) and the thermal-equilibrium initial state, has been simulated previously [14]. The results of our simulation coincide with the dependences obtained in Ref. [14].

The experimental goal is to make the diagonal part of the density matrix as close as possible to the matrix with only two elements: $|u\rangle\langle u| - |d\rangle\langle d|$. The off-diagonal elements, if there are any, can be easily eliminated by a gradient pulse or phase cycling. Fig. 3 simultaneously displays time dependences of the 6Q intensity for the thermal-equilibrium initial state; and the total 0Q intensity and a sum of intensities of two elements, $|u><u|$ and $|d><d|$, for the 6Q initial condition. One can see that at the evolution times when the 6Q intensity reaches one of its major maxima, the intensity of the two diagonal states coincides with the total 0Q intensity, i.e. no other elements are created on a diagonal of the deviation density matrix. The evolution time which has been used in the experimental realization, $0.969/D_{12}$, is shown by the arrow in Fig. 3. This time, obtained by optimizing the experimental performance, is very close to the time $0.973/D_{12}$, when the three simulated dependences in Fig. 3 simultaneously reach their maxima. The overall efficiency of converting the thermal equilibrium state into the state $|u><u|-|d><d|$, which is the product of the efficiency of converting the equilibrium state into 6Q (14%) and that of changing 6Q into $|u><u|-|d><d|$ (72%), is in a perfect agreement with the experimental value of 11%.

## 4. DISCUSSION

As seen in Fig. 2, for the pseudopure ground state, we observed an increase in intensity of the individual peak by a factor of 1.8 compared to the thermal equilibrium spectrum. On the other hand, the peak intensity for the pseudopure state was close but smaller than the corresponding peak for the equilibrium state in Ref. [5], in which the HOMQ coherence was originated from a single-spin magnetization. In our case, magnetization of all spins contributes to the HOMQ coherence. Since it has been known that selective excitation is more efficient to excite multiple quantum coherence of specific order [8], even higher increase in intensity may be achievable. Still, it should be noticed that the total signal intensity decreases with increasing number of qubits in NMR QIP with thermal ensemble [15].

As mentioned above, there exist no more than *N* peaks in the spectrum of the pseudopure ground state. The spectrum can be well-resolved even for very large clusters of coupled spins. Starting with this simplified spectrum, it is possible to apply selective pulses for manipulating individual states and observing their dynamics. As an example, relaxation and decoherence of individual quantum states can be measured with our method, and such experiments are in progress.

With increasing number of spins, it might become unpractical to distinguish all of the qubits in frequency domain. On the other hand, electron or nuclear spins in solids are expected to be good candidates for qubits, which implies that NMR method for thermal ensemble, described in this work, can be useful. In fact, there have been an approach to quantum computing without local control of qubits [16-18].

High efficiency of the proposed scheme makes it possible to increase a size of spin systems with individually addressable quantum states. We believe that this technique can be applied to larger spin clusters with truly unresolved conventional spectra and will help to experimentally explore quantum decoherence and relaxation of individual quantum states in such systems.

## ACKNOWLEDGEMENTS

This work was supported by Kent State University and US - Israel Binational Science Foundation.

**FIGURE CAPTIONS**

Fig. 1. Scheme of the pulse sequence for creating the pseudopure ground state.

Fig. 2. (a) Conventional $^1$H NMR spectrum of benzene in liquid crystal; (b) linear-response spectrum for the state $|u><u|-|d><d|$; (c) linear-response spectrum for the pseudopure ground state state $|u><u|$; flip angle of the reading pulse was 5° for all spectra.

Fig. 3. The intensity of the 6Q coherence (thick line) for the thermal-equilibrium initial state; the total intensity of 0Q coherence (dashed line) and the sum of intensities of the two diagonal elements $|u><u|$ and $|d><d|$ (thin line) for the 6Q initial state. Time is in unit of $1/D_{12}$

Fig. 1

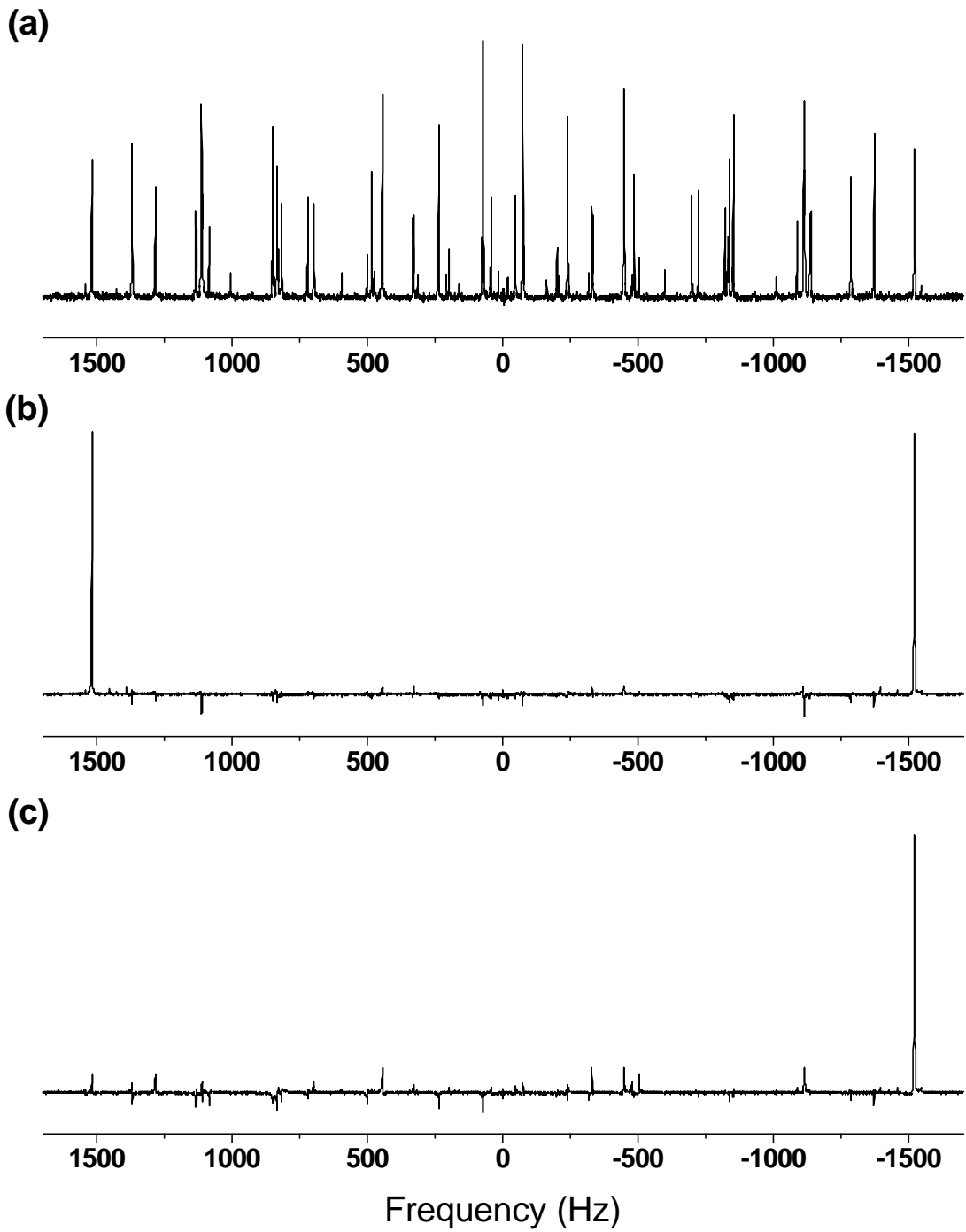

Fig. 2

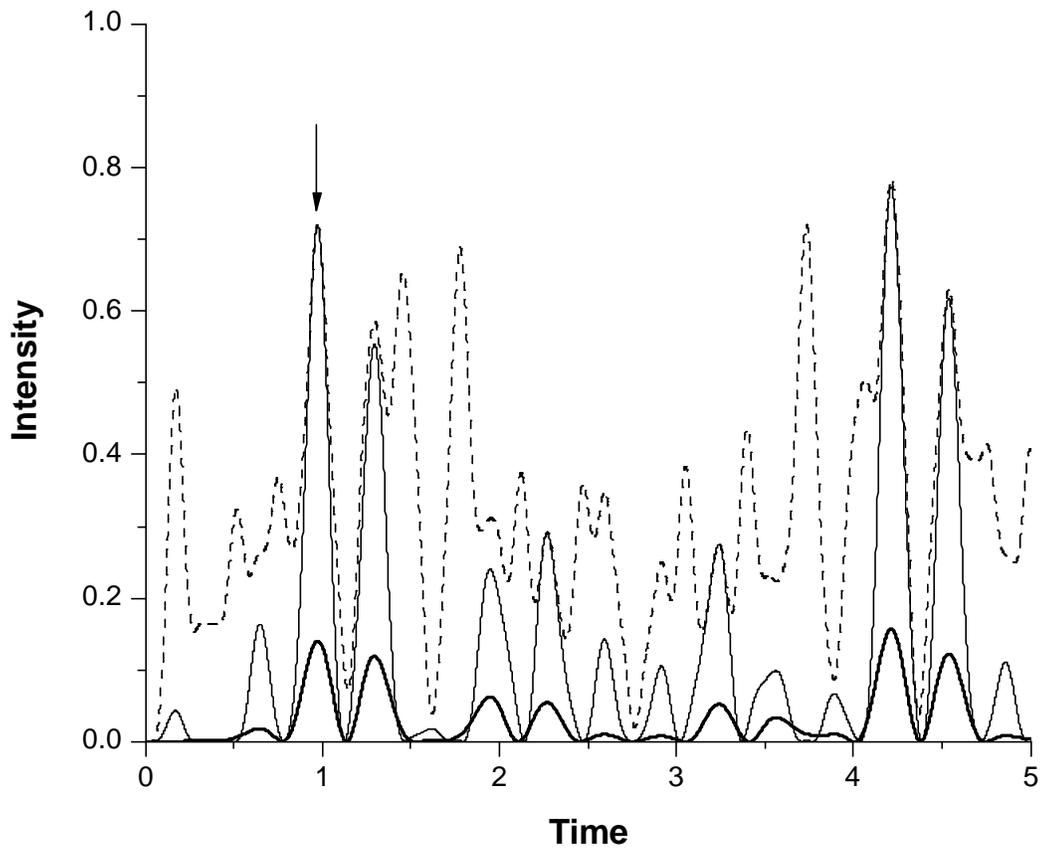

Fig. 3